\documentclass{PoS}
\usepackage{graphicx}
\usepackage[numbers,sort&compress]{natbib}
\usepackage[numbers]{natbib}

\title{Parsec-scale Nuclear Radio Structures in Seyfert Galaxies}
\ShortTitle{Parsec-scale Emission in Seyferts}
\author{\speaker{Preeti Kharb}\\
National Centre for Radio Astrophysics - Tata Institute of Fundamental Research (NCRA-TIFR), S. P. Pune University Campus, Post Bag~3, Ganeshkhind, Pune 411007, India \\ E-mail: \email{kharb@ncra.tifr.res.in}}

\abstract{Radio outflows of extents ranging from a few parsecs to a few kiloparsecs are present in Seyfert and LINER galaxies that make up the ``radio-quiet'' AGN class. AGN jets and / or starburst superwinds have been suggested to produce these outflows. We present a brief review of radio outflows in Seyfert and LINER galaxies on different spatial scales. Very long baseline interferometry (VLBI) observations of several individual Seyferts and LINERs suggest a link between AGN jets on parsec-scales and their kiloparsec-scale radio structures (KSRs). The whole range of misalignment angles present between the parsec-scale and the kpc-scale outflows in Seyfert galaxies and LINERs, supports the prevalence of bent outflows in them. Episodic AGN activity is suggested by the presence of multiple misaligned KSRs in several Seyfert galaxies in total and polarized intensity images; this latter result provides further support for an AGN jet origin of the KSRs present in Seyfert and LINER galaxies.}

\FullConference{Revisiting narrow-line Seyfert 1 galaxies and their place in the Universe - NLS1 Padova\\
9-13 April 2018 \\ Padova Botanical Garden, Italy}

\begin{document}
\section{Introduction}
Radio outflows of extents ranging from tens to hundreds of parsecs on one end, and ten to twenty kiloparsecs (kpc) on the other, have been detected in Seyfert and LINER galaxies \citep{Ulvestad81,Baum93,Thean00,Gallimore06,Kharb16,Baldi18}. The origin of these radio outflows is not unequivocally clear. \cite{Condon82} and \cite{Baum93} have suggested that these outflows could be  powered by starburst superwinds, while \cite{Colbert96} have pointed out to the distinct morphological differences between Seyfert outflows and those in starburst galaxies, supporting an AGN origin for the former. \cite{Gallimore06} have concluded from a VLA study of a complete sample that most kiloparsec-scale radio structures (KSRs) in Seyferts and LINERs are AGN-driven, but the starburst superwind contribution cannot be ruled out. It is worth noting here that while Narrow-line Seyfert 1 galaxies (NLS1s) differ from regular Seyferts in their emission-line and (possibly) black hole properties, they exhibit KSRs similar to the ones observed in Seyferts and LINERs \citep{Richards15,Berton18}.

One of the best ways to probe the AGN role is by studying the radio outflows in Seyferts and LINERs on parsec-scales via the technique of Very Long Baseline Interferometry (VLBI). VLBI arrays have angular resolutions ranging from a few milli-arcseconds (mas, e.g., VLBA\footnote{Very Long Baseline Array}) to tens and hundreds of mas (e.g., MERLIN\footnote{Multi-Element Radio Linked Interfermeter Network}). For the nearby galaxies, this translates to spatial scales of a few parsecs to a few sub-kiloparsecs (sub-kpc), respectively. Continuous tracing and accounting for the radio contribution going from parsec to sub-kpc to kpc-scales, can go a long way in settling the AGN versus the starburst superwind debate for Seyfert and LINER outflows.

VLBI observations of several Seyfert galaxies have revealed the presence of weak radio cores and one-sided or two-sided radio jets in them \citep{Mundell00,Orienti10}. The brightness temperatures ($T_b$) of the radio cores are typically of the order of $10^6-10^{11}$~K; they exhibit flat or inverted spectral indices \citep{Ho08}. However, steep spectrum radio ``cores" have also been reported in several Seyfert galaxies \citep{Roy00,Kharb10,Bontempi12}. These ``cores'' could be contaminated by the presence of steep spectrum jet emission. Alternately, the real radio cores may have failed detection at the observing frequencies and could show up at higher radio frequencies. Overall, the VLBI results are consistent with Seyferts and LINERs harbouring low luminosity AGN \citep{Falcke00,Middelberg07,Panessa13}. 

\subsection{Probing the Parsec-scale - Kiloparsec-scale Radio Connection}
VLBI jets in several Seyfert galaxies reveal bends and wiggles \citep{Roy00,Lal04,Hada13}. Jet precession has been invoked to explain these morphological peculiarties \citep[e.g.,][]{Veilleux93,Middelberg05}. Jet precession in turn could arise due to accretion disk warping, jet instabilities, or the presence of binary black holes. Mrk\,6 is one such Seyfert galaxy showing an S-shaped radio jet in MERLIN observations \citep{Kukula96}. It is a peculiar galaxy in that it shows two sets of KSRs aligned nearly perpendicular to each other \citep{Kharb06}. Our two-frequency (1.6, 5 GHz) VLBA observations of Mrk\,6 detected an inverted spectrum radio core at the higher frequency for the first time in this source, and resolved the knots in the S-shaped MERLIN jet into elongated jet-like features \citep{Kharb14a}. The precessing jet model of \cite{Hjellming81} could fit the parsec-scale emission as well as the bright edges of the north-south oriented KSR. Two episodes of AGN jet activity with precessing jets could explain the entire complex structure observed in Mrk\,6. 

A precessing radio jet could also explain the radio emission on parsec-, sub-kpc- and kpc-scales in the Seyfert / LINER galaxy with an ongoing nuclear starburst, NGC\,6764 \citep{Kharb10}. Interestingly, the precession model best-fit values of jet inclination ($\sim18^\circ$) and jet speed ($\sim0.028c$) could explain the observed jet-to-counterjet surface brightness ratio ($R_J\sim1.2$) in NGC\,6764. We examined the starburst-wind contribution to the radio emission in NGC\,6764 and another Seyfert + starburst composite galaxy, NGC\,3079 \citep[see][]{Irwin03}. We found that only about $25-30\%$ of the total radio flux density appears to arise in clear equatorial emission that could be attributable to stellar winds, in both these sources. VLBI observations of NGC\,3079 have revealed multiple misaligned jet-like features, which are not all consistently along one position angle with respect to the KSR; this lead  \cite{Kondratko05} to invoke a wide-angle parsec-scale outflow in NGC\,3079 \citep[see also][]{Mukherjee18}. In Mrk\,6 and NGC\,6764 and possibly NGC\,3079, precessing jets launched from their black hole $-$ accretion disk systems, could be powering the KSRs and dissipating on sub-kpc or kpc-scales. This could also be the case in the Seyfert galaxies NGC\,1320 and NGC\,2992: newly acquired eMERLIN data on these sources reveal $\sim20-30$ parsec-scale core-jet structures in them, that are misaligned to their KSRs (Kharb et al. 2018a, in preparation). Jet precession has also been invoked from direct multi-epoch VLBI observations in the Seyfert / LINER galaxies III Zw 2 and M81 by \cite{Brunthaler05} and \cite{Marti11}, respectively.

\begin{figure}[h]
\centerline{\includegraphics[width=7.4cm,angle=90, trim = 10 30 50 30]{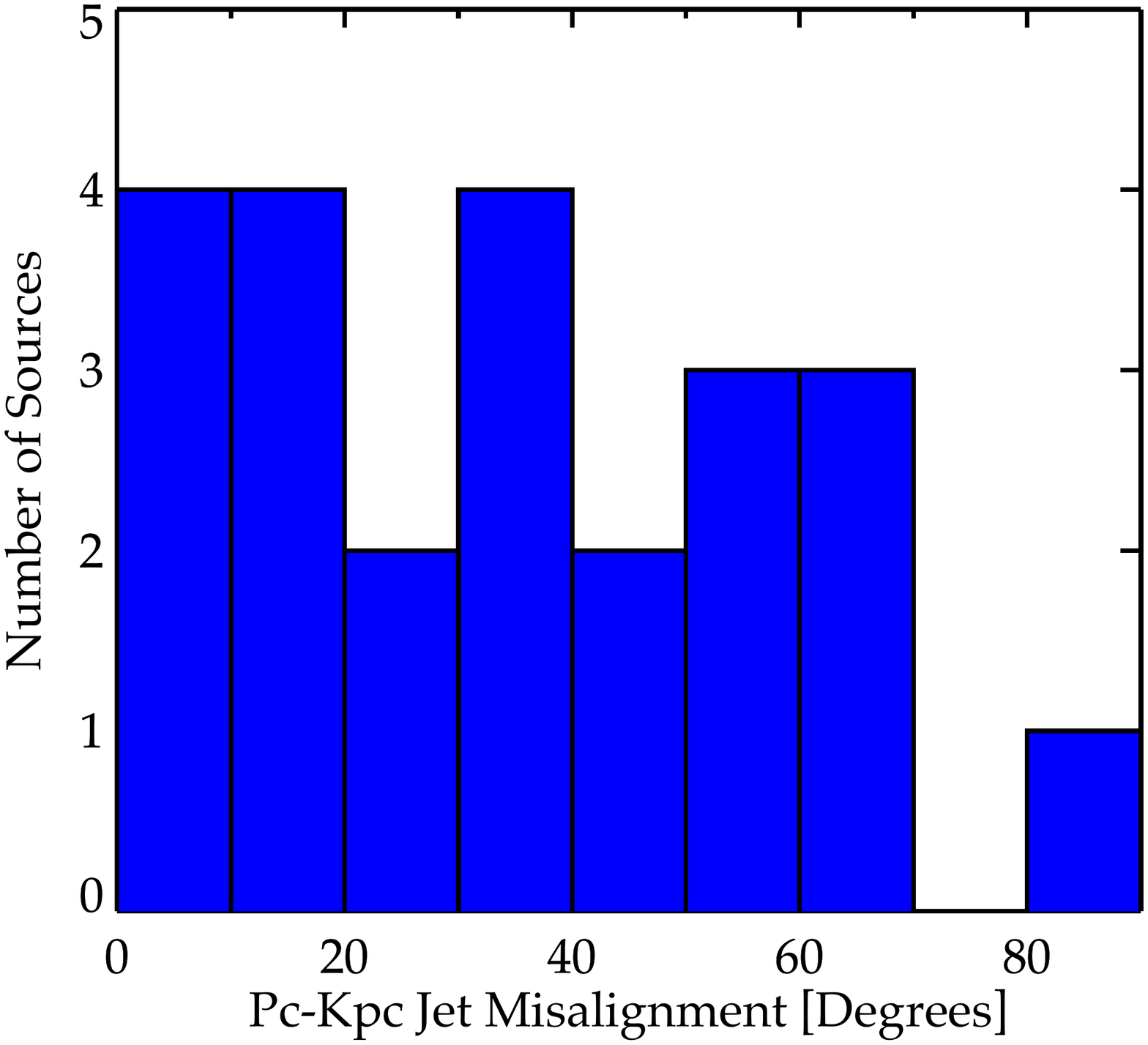}}
\caption{Jet misalignment angles from parsec to kiloparsec scales in Seyfert and LINER galaxies.}
\label{fig1}
\end{figure}
In Figure~\ref{fig1} we have plotted the parsec-to-kpc-scale jet misalignment angles in Mrk\,6, NGC\,6764, NGC\,1320, and NGC\,2992, along with others from the literature \citep{Baum93,Middelberg05}. While there could be large uncertainties in deriving the KSR position angles due to its diffuse structure (of the order of $10^\circ-20^\circ$), we note that the distribution of misalignment angles in Seyferts does not show any clear preference,  unlike what is observed in blazars which show a bimodal distribution in misalignment angles \citep{Conway93,Kharb10}. This latter result has been attributed to the presence of low-pitch helical jets in a sub-population of blazars by \cite{Conway93}. Seyfert outflows on the other hand, show the entire range of misalignment angles. Could continuously bending outflows, that are observationally sampled on different spatial scales in different sources, produce this misalignment angle distribution ? Clearly, many more Seyfert and LINER galaxies need to be examined on several different spatial scales to answer this question.

Changes in the jet propagation direction from parsec to kpc-scales have often been observed in jetted AGN, as has episodic AGN activity \citep[e.g.,][]{Saikia09}. Assuming that jet re-orientation is more likely than re-orientation of large-scale galactic disks producing starburst superwinds, the existence of multiple misaligned KSRs would support an AGN origin for them. Interestingly, secondary misaligned KSRs have been revealed in polarized emission in the Seyfert galaxies NGC\,2992 and NGC\,3079 in the Continuum Halos in Nearby Galaxies - an EVLA Survey (CHANG-ES) by \cite{Irwin17}. These secondary KSRs have not been observed in total intensity images, unlike the case of Mrk\,6. Nevertheless, they  suggest that episodic AGN activity may be a common phenomena in Seyfert and LINER galaxies.

\subsection{Could Jet Precession Point to the Presence of Binary Black Holes?}
About 1\% of SDSS AGN show double-peaked emission lines in their optical spectra \citep{Wang09}. The presence of double-peaked narrow emission lines in these double-peaked AGN (DPAGN) has been suggested to arise due to (i) binary black holes with associated narrow-line regions (NLRs), (i) jet-NLR interaction or (iii) disky NLRs. In order to search for binary black holes in Seyferts, we have been observing Seyfert DPAGN with phase-referenced VLBI. Dual frequency VLBA observations have detected a single weak $\sim$0.7 mJy radio core of size ($8\times6$~pc) in the Seyfert DPAGN KISSR\,1494 \citep{Kharb15b}. The brightness temperature of this core (T$_b\sim1.4\times10^7$~K) and steep radio spectrum ($\alpha\leq-1.5\pm0.5$;  $S_\nu\propto\nu^\alpha$), are consistent with optically thin synchrotron emission. However, the core is not ``compact'' and its brightness is not centrally concentrated; it gets resolved out in images with different weighting schemes. And yet it does not resemble a typical parsec-scale jet component. We have suggested this component to be the base of a tenuous synchrotron-emitting coronal wind. This could explain the double-peaked narrow lines as also originating in a wind-driven NLR. 

Similar dual-frequency phase-referenced VLBA observations of the Seyfert DPAGN KISSR\,1219, have revealed a one-sided $\sim70$ parsec radio jet with a steep radio spectrum \citep[$\alpha\leq-1.0\pm0.2$;][]{Kharb17a}. Using the jet-to-counterjet surface brightness ratio in KISSR\,1219 and an orientation angle consistent with its Seyfert type 2 classification ($\theta\gtrsim50^\circ$ based on the typical dusty torus half opening angle), we concluded that the jet speed had decreased from $\gtrsim0.55c$ on parsec-scales to $\gtrsim0.25c$ on kpc-scales. The radio jet was likely pushing the NLR clouds in opposite directions, giving rise to the double-peaked emission lines. This may have resulted in jet deceleration and eventual dissipation.

New VLBA observations of the Seyfert DPAGN KISSR\,434 have revealed an intriguing $\sim150$~parsec C-shaped curved jet, again suggesting jet-NLR interaction as the source of double-peaked narrow emission lines (Kharb et al. 2018b, in preparation). However, the curved jet itself could arise due to precession in a binary black hole system \citep[e.g., see][]{Rubinur17,Rubinur18}. Interestingly, we have found suggestions of a binary black hole system through multi-frequency VLBA observations of the Seyfert / luminous infrared galaxy (LIRG) NGC\,7674 \citep{Kharb17b}. This galaxy possesses a Z-shaped $\sim$0.7~kpc radio jet \citep{Momjian03}. The dual VLBA cores have T$_b\sim2 - 6\times10^7$~K and inverted spectral indices, consistent with being the synchrotron self-absorbed bases of two jets launched from two accreting supermassive black holes. Their projected separation of $\sim$0.65 mas (=0.35 parsec) makes them the closest separation binary black hole pair to be imaged with VLBI. 

\section{Summary}
Radio outflows are frequently observed in Seyfert and LINER galaxies, inspite of their ``radio-quiet" AGN status. These can span extents ranging from tens of parsecs to $10-20$ kpc or more. Nuclear starburst wind contributions to the radio emission cannot be completely ruled out, in at least some Seyferts and LINERs. However, sensitive or phase-referenced VLBI observations that can probe faint parsec-scale radio emission, more often than not reveal weak radio cores and wiggly radio jets in Seyfert and LINER galaxies. In individual sources with multi-scale data on parsec, sub-kpc and kpc-scales, an AGN jet, which could be curved in many cases, can connect the emission on different spatial scales, making the case for an AGN origin for the radio outflows. Curved jets could suggest jet-ISM interaction or precession, which in turn could suggest the presence of binary black holes or accretion disk instabilities. 

\section*{Acknowledgements}
We thank the anonymous referee for a positive response to our paper.
This conference has been organized with the support of the Department of Physics and Astronomy ``Galileo Galilei'', the University of Padova, the National Institute of Astrophysics INAF, the Padova Planetarium, and the RadioNet consortium. RadioNet has received funding from the European Union's Horizon 2020 research and innovation programme under grant agreement No~730562. 

\setlength{\bibsep}{0pt}
\bibliographystyle{JHEP}
\bibliography{ms}

\end{document}